\documentstyle[12pt]{article}
\begin{document}
\title{LARGE SCALE STRUCTURES IN THE UNIVERSE}
\author{B.G. Sidharth$^*$\\
Centre for Applicable Mathematics \& Computer Sciences\\
B.M. Birla Science Centre, Adarsh Nagar, Hyderabad - 500063 (India)}
\date{}
\maketitle
\footnotetext{\noindent E-mail:birlasc@hd1.vsnl.net.in}
\begin{abstract}
In this brief communication we show why superclusters would naturally arise
in the universe.
\end{abstract}
It is well known in the theory of the Random Walk or Brownian motion that
\begin{equation}
R = \sqrt{N}l\label{e1}
\end{equation}
holds where $R$ denotes the dimension of the system, $N$ the number of steps
or events and $l$ represents a mean free path\cite{r1}. It has already been
argued\cite{r2,r3} that in the context of the universe as a whole,
$N$ representing the total number of particles, (\ref{e1}) gives the
Eddington relation with $l$ being the pion Compton wavelength.\\
Let us now consider $N \sim 10^6$ constituents in the universe. Then (\ref{e1})
gives
\begin{equation}
l \sim 10^{25} cm\label{e2}
\end{equation}
Indeed there is observational evidence for (\ref{e2}):
We can easily see that (\ref{e2}) holds for superclusters, both in terms of
their size $l$ and their number $N$\cite{r4}. Further, (\ref{e1}) shows
that these superclusters must have a two dimension character, which is also
true.\\
It is ofcourse well known that one cannot apply the theory of Brownian motion
to stars or even galaxies because they are gravitationally bound. However
for superclusters with the huge separating voids, Brownian motion would be
a reasonable approximation, as can be seen by the fact that (\ref{e2}) is
valid. So it is natural that such superclusters should arise.\\
It is interesting to note also that recently, in a completely different
context, it was suggested\cite{r5} that there could be a large scale
quantization, giving precisely (\ref{e2}) as the quantized length.

\end{document}